\RequirePackage{amsmath}
\RequirePackage{amssymb}
\documentclass[runningheads]{llncs}

\usepackage[binary-units]{siunitx}

\usepackage{hyperref}

\usepackage[caption=false]{subfig}

\usepackage{graphicx}

\usepackage{tikz}
\usepackage{pgfplots}

\begin{document}
\title{Coloured Ring Confidential Transactions\thanks{This work was partially funded by the Federal Ministry of Economic Affairs and
Energy on the basis of a decision by the German Bundestag. The authors thank Henning Kopp for the technical discussions  during the sketching phase. Preprint accepted at ESORICS '18 CBT workshop. The final publication is available
at \href{https://link.springer.com}{link.springer.com}.
}}
%
%
\author{Felix Engelmann \and
Frank Kargl\and
Christoph B\"osch}
\authorrunning{F. Engelmann et al.}
%
\institute{Institute for Distributed Systems, 
Ulm University, 
Ulm, Germany\\
\email{\{felix.engelmann,frank.kargl,christoph.boesch\}@uni-ulm.de}}
\maketitle              
\begin{abstract}
Privacy in block-chains is considered second to functionality, but a vital requirement for many new applications, e.g., in the industrial environment.
We propose a novel transaction type, which enables privacy preserving trading of independent assets on a common block-chain.
This is achieved by extending the ring confidential transaction with an additional commitment to a colour and a publicly verifiable proof of conservation.
With our coloured confidential ring signatures, new token types can be introduced and transferred by any participant using the same sized anonymity set as single-token privacy aware block-chains. 
Thereby, our system facilitates tracking assets on an immutable ledger without compromising the confidentiality of transactions.

\end{abstract}
\section{Introduction}

Trading is a basic human trait that extends to the digital world.   Individual trading without the need of intermediaries is enabled by block-chain technology.   Participants of a block-chain reach a global consensus on which trades are valid and in which order.   To achieve this, all transactions must be validated by peers and checked for violations of conservation rules, e.g., creating an asset out of thin air.   The basic approach is to use plain-text transaction receipts, visible for everyone which makes validation of the transactions straightforward.

The issue with these plain-text receipts is, that trades often include valuable information for other parties, using the knowledge for their leverage.   Independent research~\cite{ringct,zcash} realised, that privacy in block-chain systems is important to support the same features as analogue trades.   Monero therefore introduced ring confidential transactions to hide the sender identity (using ring signatures), the recipient identity (using  one-time payment addresses), and the amount transferred (using  commitments) from the public, while maintaining the possibility to verify the conservation.  The real sender is indistinguishably concealed within a set of decoys.   To prove ownership of an asset, which is attached to a public key, a ring signature is used instead of a regular digital signature.   The verifier of a ring signature can check that the signer knows at least one of the corresponding private keys, but not which one.   To prevent double spending of the same asset, the ring signature has to include a specific tag, which stores the identity of the signer in an encrypted form.   If two signatures have the same tag, they can be linked and the second signature is invalid.   Thus, no two assets can belong to the same public key.
This requirement demands for one-time recipient addresses, which, in addition, serve the purpose of hiding the recipient from the public.   Therefore, a one-time key is derived from the long-term recipient public key, for which only the recipient can derive the correct one-time private key.   These one-time addresses prevent multiple transaction outputs to be linked to a common owner.   This works well, if all transaction inputs are of equal value.   However, transaction inputs with different values, still allow for deducing the real sender by comparing the transaction in- and outputs.   In order to prevent this kind of sender derivation and to add privacy, the transaction value is hidden inside a commitment.   Additively homomorphic Pedersen commitments~\cite{pedersen} allow the sender to prove that the input minus the output of a transaction is zero, thereby proving the conservation without disclosing the amount. A detailed description of the techniques used is summarised by Alonso et al.~\cite{moneroprivacy}.

The added privacy compared to fully visible transaction receipts restricts features, such as Turing complete smart contracts, which are common on non-privacy aware block-chains.   Smart contracts are recipients, whose behaviour is governed by code.   A prominent use-case of smart contracts is the management of tokens.   These tokens can be sub-currencies or used to track assets independent of the block-chain's native currency.

In this paper, we introduce an extension to the ring confidential transaction to support sub-currencies with the benefit of privacy aware trading.   Our construction features multiple coexisting asset types, also known as colours.   A transaction can transfer exactly one colour, but the decoy inputs can be from any colour, having the same anonymity set as single-colour privacy aware block-chains.   The colour of the transaction is only known to the interacting parties ( sender and recipient of the current transaction), but not to anyone else, achieving a fully privacy aware verification of colour conservation from inputs to outputs.

With the help of our new transaction type, all participants of the block-chain can introduce new token types for their own purposes.   The consensus verifies that a new colour does not yet exist to prevent unauthorised issuance of existing tokens.   All the new tokens will benefit from the privacy aware transactions without the barrier of creating an independent chain per colour.   A new block-chain per colour reduces the opportunities for decoys in a transaction which negatively impacts the privacy of the whole system.   On top, multiple colours on a single block-chain facilitate future on-chain atomic swap operations between colours.

\section{Preliminaries}
Our contribution extends the ring confidential transaction (ringCT), which is prominently used in Monero for the \texttt{RCTTypeFull}, to support a colour attribute for in- and outputs of a transaction.
In this section, we describe the required building blocks, an additively homomorphic commitment scheme and a linkable ring signature, which are the same as for the ringCT.
In addition to this, a full ringCT requires range proofs, but as our extension does not require an adaptation thereof, we refer the reader to the work of Noether et al.~\cite{ringct} for the full construction.


We use elliptic curve cryptography for our commitments and signatures.   An elliptic curve is a group with the possibility to add an element, also called point, to another or itself resulting in a new point on the curve.   This allows for the multiplication of a scalar $x$ to a point.   A curve standardises a base point $G$, which is the generator of a preferably large subgroup.   Elliptic curves are suited for cryptography, as calculating $x$ given $P=xG$ is hard, known as the discrete logarithm problem.   This property can be used to generate private-public key pairs $(\text{sk}=x,\text{pk}=xG)$.

\subsection{Pedersen commitments}
To hide a value $a$ in an Elliptic Curve Pedersen commitment~\cite{pedersen} requires two points, where one can be the base point $G$ and the second point $H=\psi G$ must be created, such that $\psi$ is unknown to anyone. A \emph{nothing up my sleve} generation of $H$ can be generated by hashing the base point with a hash function $\mathcal{H}$ mapping from a point to another point with $H=\mathcal{H}(G)$. $G$ and $H$ are the public parameters of the given system.
To build a commitment to a value $a$, a secret, random blinding factor $x$ is generated and then combined to
\begin{equation*}
 C(a,x) = xG+aH.
\end{equation*}
Pedersen commitments are perfectly hiding and computationally binding under the discrete logarithm assumption. A commitment $C(0,x)=xG$ is binding but not hiding.
By publishing the point $C(a,x)$, the sender commits to the value $a$, and can only change the choice by brute-force searching for a different pair $x',a'$ satisfying 
$
C(a,x)=xG+aH = x'G+a'H 
$
which has a negligible chance of success.

The Pederson commitment has the desirable property, that finding the value $y$ with $C(a,x)=yG$ for $a\not=0$ and $x\not=0$ is difficult according to the discrete logarithm problem. However, given that $a=0$, the commitment is reduced to
$
 C(0,x)=xG+0H=xG.
$
Then the private key to the committed point is $x$ which is used to sign the commitment and thus proving knowledge of $x$.

An additional feature of the commitments are their homomorphicity in regard to addition.
Three commitments $C_1(5,x_1), C_2(3,x_2), C_3(2,x_3)$ can be summed together like
$
 C_1-(C_2+C_3)=x_1G+5H-(x_2G+3H+x_3G+2H)=(x_1-x_2-x_3)G
$
resulting in a commitment to zero $C_0$ with secret key $x_1-x_2-x_3$.
Whoever can sign the sum of the commitments proves knowledge of all the components and proves that the sum of values is $0$.
For values which should be in plain-text, but which are needed to perform calculations, a commitment can be opened, by immediately disclosing $a$ and $x$.

\subsection{Multilayered Linkable Spontaneous Ad-hoc Group Signature}
\label{sec:mlsag}
The second building block we require for our colour extension is the Multilayered Linkable Spontaneous Ad-Hoc Group Signature (MLSAG). This is a modification of the Fujisaki-Suzuki (FS)~\cite{fs} and the Liu-Wei-Wong (LWW)~\cite{lww} signatures to increase their space efficiency.
It provides a signature where the signer can prove knowledge of a set of private keys which are embedded in a larger set of decoys. The verifier can not deduce for which subset the signer knows the private keys. If any one of the private keys is reused, the two resulting signatures can be linked together, preventing double spending of a single output.
The keygen, sign, verify and link algorithms are described according to Noether et al.~\cite{ringct}.

\begin{description}
 \item[$(P^j,x_j)\leftarrow$ ML.Keygen$(1^\lambda)$:] Generate a vector of $m$ private keys $x_{i}$ for $i=1,\dots,m$ with the corresponding public keys $P^i=x_iG$.
 \item[$(P_i^j,I_j)\leftarrow$ML.Keyselect$(P^j)$:] Select a set of $n-1$ vectors, each containing $m$ public keys $\{P_i^j\}_{i=1,\dots,n}^{j=1,\dots,m}$ from other users. For a secret index $\pi$, corresponding to the signer, all the secret keys $x_j$ must be known, such that $x_jG=P_\pi^j$ and let $I_j=x_j\mathcal{H}(P_\pi^j)$ for $j=1,\dots,m$.
 
\item[$\sigma\leftarrow$ML.Sign$(\mathfrak{m},P_i^j,x_j,I_j)$:] Let $\mathfrak{m}$ be the message to sign. For $j=1,\dots,m$ and $i=1,\dots,\allowbreak{\pi-1},\allowbreak{\pi+1},\dots,n$ draw $s_i^j$ and $\alpha_j$ as secure, random scalars. With a hash function $\mathfrak{h}:\{0,1\}^*\to \mathbb{Z}_q$, compute
$
 L_\pi^j=\alpha_jG $ and $ R_\pi^j=\alpha_j\mathcal{H}(P_\pi^j)
$ for $j=1,\dots,m$. Continue with the vector $i=\pi+1$ as
\begin{align*}
  c_{\pi+1}&=\mathfrak{h}(\mathfrak{m},L_\pi^1,R_\pi^1,\dots,L_\pi^m,R_\pi^m)\\
  L_{\pi+1}^j&=s_{\pi+1}^jG + c_{\pi+1}P_{\pi+1}^j \text{ and } R_{\pi+1}^j=s_{\pi+1}^j\mathcal{H}(P_{\pi+1}^j)+c_{\pi+1}I_j
\end{align*}
and calculate $L$ and $R$ for each increment of $i\mod n$ until $i=\pi-1$ like
\begin{align*}
  c_{\pi-1}&=\mathfrak{h}(\mathfrak{m},L_{\pi-2}^1,R_{\pi-2}^1,\dots,L_{\pi-2}^m,R_{\pi-2}^m)\\
  L_{\pi-1}^j&=s_{\pi-1}^jG + c_{\pi-1}P_{\pi-1}^j \text{ and } R_{\pi-1}^j=s_{\pi-1}^j\mathcal{H}(P_{\pi-1}^j)+c_{\pi-1}I_j .
\end{align*}

Given $c_\pi=\mathfrak{h}(\mathfrak{m},L_{\pi-1}^1,R_{\pi-1}^1,\dots,L_{\pi-1}^m,R_{\pi-1}^m)$, we calculate $s_\pi^j$ with
$
 \alpha_j=s_\pi^j+c_\pi x_j \mod l
$
 (modulus curve order $l$) and the output consists of
\begin{equation}
 \sigma=(c_1,s_1^1,\dots,s_1^m,s_2^1,\dots,s_2^m,\dots,s_n^1,\dots,s_n^m,I_1,\dots,I_m).
 \label{eq:signature}
\end{equation}

\item[$0/1 \leftarrow$ML.Verify$(\mathfrak{m},\sigma,P_i^j)$:]
Starting with $i=1$ and $c_1$, calculate $L_i^j$ and $R_i^j$ for all $i$ and $j$. If $c_{n+1}=c_1$, the signature is valid and 1 is returned, 0 otherwise.
\item[$0/1 \leftarrow$ML.Link$(\sigma,\sigma')$:] If the signatures $\sigma$ and $\sigma'$ share an $I_j$, they used the same private key $x_j$ in the signing process and 1 is returned, 0 otherwise.
\end{description}

\noindent The MLSAG signature scheme must satisfy the the following correctness conditions: For every $\lambda,m\in\mathbb{N}$, every $n\in\mathbb{N}\backslash\{1\}$, every $(P^j,x_j)\leftarrow\mathbf{ML.Keygen}(1^\lambda)$, and every $\mathfrak{m}$, it holds with high probability that
\begin{equation*}
 \mathbf{ML.Verify}(\mathfrak{m},\sigma\leftarrow\mathbf{ML.Sign}(\mathfrak{m},P_i^j\leftarrow\mathbf{ML.Keyselect}(P^j),x_j,I_j),P_i^j)=1.
\end{equation*}

\noindent The MLSAG satisfies the following security properties which are proven in the original LWW signature description~\cite{lww} and the construction by Noether et al.~\cite{ringct}: 
 \begin{itemize}
  \item \emph{Unforgeability:} negligible probability of producing a valid signature without knowledge of all private keys in one vector.
  \item \emph{Linkability:} negligible probability of being able to produce two different signatures using the same private key in both.
  \item \emph{Signer Ambiguity:} negligible additional probability of guessing the secret index, even by knowing private keys of decoy inputs.
 \end{itemize}

\section{Our Coloured Ring Confidential Transaction}
Having explained the necessary building blocks, we proceed with a detailed description of the \texttt{RCTTypeFull} ringCT and highlighted the additional elements required for our extension in {\color{red}red}.

\subsubsection{In- and Outputs.}
Our transaction requires inputs, which are outputs of previous transactions.
The sender selects $m$ inputs to be used.
Depending on how many decoys per input ($n-1$) the sender wants to include in the transaction, additional $m\cdot(n-1)$ inputs are selected. 
Each input contains a public key $P_i^j$, which was generated as a one-time payment address.
The amount $a$ each input holds is stored in a Pedersen commitment $C_i^j(a,b)$ with blinding factors $b$. The sender only knows $a_{j,in}$ and $b_{j,in}$ and $x_j$ for the inputs $(P_\pi^j=x_jG,C_\pi^j(a_{j,in},b_{j,in}))$ under its control.
All real inputs make up one vector 
\begin{equation*}
\{(P_{\pi}^1,C_\pi^1(a_{1,in},b_{1,in})),\dots,(P_\pi^m,C_\pi^m(a_{m,in},b_{m,in})\} 
\end{equation*}
at the secret index $\pi$.
The decoy vectors at $i=1,\dots,\pi-1,\pi+1,\dots,n$ are assembled equally, with neither knowledge of the private keys for $P_i^j$ nor of the blinding factors and amounts of the commitments $C_i^j$.
 
We introduce the colour property as an additional commitment in each input.
Colours are defined as scalars $f_i$. Each input gets an additional commitment $F_i^j$ to a colour. For the sender owned inputs, the colours $f_{j,in}$ and blinding factors $u_{j,in}$ of the commitments $F_\pi^j(f_{j,in},u_{j,in})$ are known. 
An input $(P_\pi^j,C_\pi^j(a_{j,in},b_{j,in}),{\color{red} F_\pi^j(f_{j,in},u_{j,in})})$ is now composed of the recipient one-time key and two commitments.
The $q$ outputs of a transaction are also represented as a tuple of three elements $(P_k,C_k(a_{k,out},b_{k,out}),{\color{red} F_k(f_{out},u_{k,out})})$ for $k=1,\dots,q$ with the blinding factors $b_{k,out}$ and $u_{k,out}$ randomly drawn and secret.

\subsubsection{Conservation.}

The sum of amounts of all inputs into a transaction must always be greater or equal to the sum of all output amounts, so the plain-text equation
$
 \sum_{j=1}^m a_{j,in}-\sum_{k=1}^q a_{k,out}=0
$
translates to a commitment equation
\begin{equation}
 \sum_{j=1}^m C_i^j-\sum_{k=1}^q C_k=C_0^i
 \label{eq:zero-commit}
\end{equation}
resulting in a commitment $C_0^i$ to zero.
For $i=\pi$ in Eq. (\ref{eq:zero-commit}), the signer knows all amounts $a_{j,in}$ and blinding factors $b_{j,in}$, which make up the private key to $C_0^\pi$.
This conservation ensures, that no asset is created in a transaction.

To ensure, that the real inputs are all from the same colour, the colour commitments $F_i^j$ are checked in pairs to the colour commitment of the first output $F_1$. Again we can use a commitment to zero
$
 F_i^j-F_1 = F_0^{i,j},
$
which does not disclose the colour. 
Unlike the summation of the amounts, comparing aggregate commitments utilising the homomorphic property is not secure and could lead to the following attack.
An attacker creates a transaction with two input colours $f_{in}-\epsilon$ and $f_{in}+\epsilon$ and an output $f_{out}$. If we only verify that
$ f_{in}-\epsilon + f_{in}+\epsilon = 2 f_{out}$
the inputs are not necessarily from the same colour.
This conservation rule only supports one colour per transaction.  
Transactions with multiple colours involved, maintaining the signer ambiguity is supported by an extended version of our scheme currently in development.

\subsubsection{Signature.}

The $n$ commitments from the amounts and $n\cdot m$ commitments from the colour checks can now be signed by an MLSAG from Section~\ref{sec:mlsag}.
To bind a zero commitment to the originating spend key, and to have independent link tags, the public key is added to the commitment. As the sender knows the private key $x_j$ to the spend key $P_\pi^j$ and the components of the commitments to colour and value, it can still sign the sum of commitment and $P_\pi^j$ with $x_{m+1+j}=x_j+f_{j,in}-f_{1,out}$.
The following set of vectors is used as key input $P_i^j$ into the $\textbf{ML.sign}(\mathfrak{m},P_i^j,x_j,I_j)$ algorithm with $I_j$ from the \textbf{ML.Keyselect} algorithm:
\begin{equation*}
\begin{split}
P:=&\Biggl[\Biggl\{P_1^1,\dots, P^m_1,\sum_{j=1}^m (P_1^j + C_1^j) - \sum_k C_k ,{\color{red} P_1^1 + F_1^1 - F_1, \dots, P^m_1 + F^m_1 - F_1}\Biggr\},\\
 \dots,&\Biggl\{P_\pi^1,\dots, P^m_\pi,\sum_{j=1}^m (P_\pi^j + C_\pi^j) - \sum_k C_k ,{\color{red} P_\pi^1 + F_\pi^1 - F_1, \dots, P^m_\pi + F^m_\pi - F_1 }\Biggr\},\\
 \dots,&\Biggl\{P_n^1,\dots, P^m_n,\sum_{j=1}^m (P_n^j + C_n^j) - \sum_k C_k ,{\color{red}P_n^1 + F_n^1 - F_1, \dots, P^m_n + F^m_n - F_1 }\Biggr\}\Biggr].
 \end{split}
 \label{eq:rcolour}
\end{equation*} 
 
\subsubsection{Output Proofs.}

The amounts are values modulus the curve order $l$, so overflows can be used to create new assets in a transaction. 
To counter this, the ringCT uses range proofs~\cite{bulletproof,borromean} to confine the output amounts to the interval $[0,2^{64}]$.
Our extension has to make sure, that all outputs are commitments to the same colour. 
We achieve this by appending $q-1$ signatures for the zero commitments
$
 F_1-F_k = F_0^k
$
for $k=2,\dots,q$.

\subsubsection{Public Verification.}

The complete transaction with the references of the inputs and outputs and the ring signature $\sigma$ is broadcast and anyone is able to verify the transaction and the conservation of assets. Therefore the vectors of public keys $P_i^j$ are read from the referenced inputs together with the amount and colour commitments. The points for checking the conservation are calculated. The transaction is accepted if \textbf{ML.Verify}$(\sigma,\mathfrak{m},P_i^j)=1$ and \textbf{ML.Link}$(\sigma,\sigma')=0$ for all other transactions $\sigma'$.

\section{Discussion}

In this section we evaluate the theoretical impact of our extension and discuss its implications on the privacy of the whole system. 

\subsubsection{Correctness. }
The correctness of the proposed scheme, is satisfied by the availability of a rightful owner of an output and it's corresponding key to transfer the funds of one colour to another address. This is given under the correctness of the non-colour aware ringCT. The restriction of real in- and outputs being of the same colour only separates the transactions into different asset types, but within each of them, funds can be transferred.

\subsubsection{Size and Performance Overhead.}

The MLSAG signature size increases significantly compared to a \texttt{RCTTypeFull} transaction.
The current MLSAG signature (Eq. (\ref{eq:signature})) requires 
$
 (n(m+1)+1 + m) 32 + \epsilon
$
Bytes, with $\epsilon$ being the size of variable length encoded positions of the ring members.
In addition to this, the $q$ outputs require
$
 q (1+64\cdot2+64) 32
$
Bytes for the Borromean range proofs~\cite{borromean} including signatures and commitments proving a range of \SI{64}{\bit}.

Our extension depends on longer vectors because of the colour equivalency proofs.
The signature size then increases by $n\cdot m$ additional random values $
 s_{1,m+2},\dots, s_{1,m+1+m}, \dots, s_{n,m+2},\dots,s_{n,m+1+m}
$ to 
$
 (n(m+1+m)+1 +m) 32 + \epsilon
$
Bytes.
The range proofs for the amount stay exactly the same.
To prove that the colours of the outputs are all the same, we need additional $q-1$ signatures for pairwise commitments to zero.

The range proofs use most of the space of the current transactions, so that our increase in signature size is quite negligible. Only for a high number of inputs, the impact is significant. 
Comparing only the signature sizes, our new approach requires approximately twice the space. With the introduction of bullet proofs~\cite{bulletproof} the range proof size will no longer increase linearly, but logarithmically leading to a greater influence of the colour overhead.

\subsubsection{Security Analysis.}

Our construction uses the MLSAG and Pedersen commitments in a unmodified version as black-boxes and can therefore rely on the guarantees provided by these primitives.

The addition of a token colour provides a second attribute with which transactions can be related.
In a transaction with multiple inputs, an attacker who knows that referenced inputs have different colours can discard these from the anonymity set. Assuming a worst-case uniform distribution over colours of transactions the probability of selecting a complete decoy vector with the same colour is
$
 \frac{1}{\chi^m}
$
and vanishing with the number of inputs $m$ and a total of $\chi$ colours. For a more likely distribution of transaction frequencies modelled by a power law, with most outputs in the native colour, the probability to find a one-colour decoy is higher.
A Zipf distribution results in a probability of approximately one in each 20 decoy vectors having 2 equal colours for a transaction with two inputs and a reasonable 200 colours in total.

\subsubsection{Initial Colour Creation.}

The ability to transfer privacy aware coloured tokens requires a token issuing protocol to begin with. A simple way is to allow an additional output in a transaction granting the output address a defined number of tokens in a new colour. With an open colour commitment, the transaction is only valid, if the colour is new with respect to all previous colour initiation transactions.
Depending on the usage of the new token type, the amount can be an open commitment, to publicly announce the total supply of the token, or, if not needed, be confidential.

\section{Related Work}


\subsubsection{Confidential Assets.}

Andrew Poelstra et al.~\cite{confidentialassets} created a protocol to hide transaction values and allow the transaction of multiple assets on the same block-chain. They use Pedersen commitments to store the amount of each UTXO and because of it's homomorphic properties, can publicly verify the conservation.
To mark different assets, a \emph{asset tag} in the form of a commitment to a curve point is added to each output. These asset tags must be generated, such that no factor in any pair of assets is known to anyone. By the discrete logarithm assumption it is hard to verify that no such factors exist for newly introduced assets. While this is no limitation of block-chains with a predefined number of different assets, the dynamic addition of new asset types by untrusted participants can introduce asset which might have a nontrivial factor to an existing one. Moreover their scheme does not support sender set anonymity.

\subsubsection{Hidden in Plain Sight: Transacting Privately on a Blockchain.}

Oleg Andreev at Chain Inc. also proposed a multi asset transaction\footnote{\url{https://blog.chain.com/hidden-in-plain-sight-transacting-privately-on-a-blockchain-835ab75c01cb}} with the same fundamental techniques as the confidential assets. They also represent different assets as orthogonal curve points and need to verify that the factors between assets are not known to anyone. As the work before, sender anonymity can not be satisfied by hiding the real transaction input in an anonymity set.

\section{Conclusion}

We introduced an extension to the ring confidential transaction to support multiple colours of tokens to coexist on one block-chain.
The transaction is publicly verifiable to transfer only assets in a single colour, without disclosing it.
To achieve a high grade of anonymity, the decoy inputs can be of any token colour.
Thereby, we allow for an easy issuance of privacy preserving tokens which benefit from each other by disguising themselves with each other.
On top, our approach can use all the existing privacy preserving mechanisms in place, such as an anonymous peer to peer network, to maintain the privacy of the participants.
Compared to contending solutions, we only require a small adaptation of an existing protocol.

%
%
%
\bibliographystyle{splncs04}  
\bibliography{bib}

\begin{thebibliography}{1}
\providecommand{\url}[1]{\texttt{#1}}
\providecommand{\urlprefix}{URL }
\providecommand{\doi}[1]{https://doi.org/#1}

\bibitem{moneroprivacy}
Alonso, K.M., Joancomartì, J.H.: Monero-privacy in the blockchain

\bibitem{bulletproof}
B{\"u}nz, B., Bootle, J., Boneh, D., Poelstra, A., Wuille, P., Maxwell, G.:
  Bulletproofs: Efficient range proofs for confidential transactions. Tech.
  rep., Cryptology ePrint Archive, Report 2017/1066, 2017.
  https://eprint.iacr.org/2017/1066 (2017)

\bibitem{fs}
Fujisaki, E., Suzuki, K.: Traceable ring signature. In: International Workshop
  on Public Key Cryptography. pp. 181--200. Springer (2007)

\bibitem{lww}
Liu, J.K., Wei, V.K., Wong, D.S.: Linkable spontaneous anonymous group
  signature for ad hoc groups. In: Australasian Conference on Information
  Security and Privacy. pp. 325--335. Springer (2004)

\bibitem{borromean}
Maxwell, G., Poelstra, A.: Borromean ring signatures (2015)

\bibitem{ringct}
Noether, S., Mackenzie, A., et~al.: Ring confidential transactions. Ledger
  \textbf{1},  1--18 (2016)

\bibitem{pedersen}
Pedersen, T.P.: {Non-Interactive and Information-Theoretic Secure Verifiable
  Secret Sharing}. In: CRYPTO'91. pp. 129--140 (1991)

\bibitem{confidentialassets}
Poelstra, A., Back, A., Friedenbach, M., Maxwell, G., Wuille, P.: Confidential
  assets. In: Financial Cryptography Bitcoin Workshop.
  https://blockstream.com/bitcoin17-final41.pdf (2017)

\bibitem{zcash}
Sasson, E.B., Chiesa, A., Garman, C., Green, M., Miers, I., Tromer, E., Virza,
  M.: Zerocash: Decentralized anonymous payments from bitcoin. In: Security and
  Privacy (SP), 2014 IEEE Symposium on. pp. 459--474. IEEE (2014)

\end{thebibliography}
%
%
%
%
%
%
%
\end{document}